\documentclass[aps,prl,twocolumn,amsmath,amssymb,superscriptaddress]{revtex4}
\usepackage[dvips]{graphicx}% Include figure files
\usepackage{dcolumn}% Align table columns on decimal point
\usepackage{bm}% bold math
\usepackage{float,color}
\usepackage{gensymb}

\begin{document}

\title{
  Ground state structure of polymeric carbon monoxide with high energy density}

\author{Kang Xia}
\affiliation{
National Laboratory of Solid State Microstructures,
School of Physics and Collaborative Innovation Center of Advanced Microstructures,
Nanjing University, Nanjing, 210093, P. R. China}

\author{Jian Sun}
\email{E-mail: jiansun@nju.edu.cn}
\affiliation{
National Laboratory of Solid State Microstructures, 
School of Physics and Collaborative Innovation Center of Advanced Microstructures, 
Nanjing University, Nanjing, 210093, P. R. China}

\author{Chris J.\ Pickard}
\affiliation{
Department of Materials Science \& Metallurgy, University of Cambridge,
27 Charles Babbage Road, Cambridge CB3 0FS, UK }
\affiliation{
 Advanced Institute for Materials Research, Tohoku University
2-1-1 Katahira, Aoba, Sendai, 980-8577, Japan }

\author{Dennis D.\ Klug}
\affiliation{
Steacie Institute for Molecular Sciences, National
Research Council of Canada, Ottawa, K1A 0R6, Canada }

\author{Richard J.\ Needs}
\affiliation{
Theory of Condensed Matter Group, Cavendish Laboratory, J
J Thomson Avenue, Cambridge CB3 0HE, UK }

\date{\today}

\begin{abstract}

  Crystal structure prediction methods and first-principles 
  calculations have been used to explore low-energy
  structures of carbon monoxide (CO).
  Contrary to the standard wisdom, the most stable structure
  of CO at ambient pressure was found to be a polymeric structure of
  $Pna2_1$ symmetry rather than a molecular solid.
  This phase is formed from six-membered (4 Carbon + 2 Oxygen) rings
  connected by C=C double bonds with two double-bonded oxygen atoms
  attached to each ring.
  Interestingly, the polymeric $Pna2_1$ phase of CO has a much higher energy density
  than trinitrotoluene (TNT).
  On compression to about 7 GPa, $Pna2_1$ is found to transform into
  another chain-like phase of $Cc$ symmetry which has similar ring
  units to $Pna2_1$.  On compression to 100 GPa it is energetically
  favorable for CO to polymerize to form a
  single-bonded $Cmcm$ phase from another structure of $Cmca$ symmetry
  composed of units similar to those found in the single-bonded
  $I2_12_12_1$ structure.
 Thermodynamic stability of these structures was verified 
 using calculations with different density functionals,
 including hybrid and van der Waals corrected functionals.

\end{abstract}

\maketitle

Carbon monoxide (CO) has the strongest known chemical bond. It has
been used extensively as a probe molecule for investigating oxidation
reactions in catalysis, and it is an important industrial gas.
 The study of
polymerization of molecular crystals
\cite{Hemley2000,Schettino2003} 
is essential for
understanding their fundamental physics and chemistry, and for
discovering new materials with useful properties such as ``high energy
density'' \cite{Lipp2005} and ``superhardness''
\cite{Yoo1999,Eremets2004}.  The phase diagram and polymerization of
CO have consequently been studied in depth over several decades
\cite{Cromer1983,Mills1984,Katz1984,Mills1986,Fracassi1986,Frapper1997,
  Lipp1998,Bernard1998,Lipp2005,Evans2006,Ceppatelli2009,Sun2011}.
Transformations from van der Waals bonded molecular
phases to covalently bonded networks have been explored in similar
systems, such as N$_2$\cite{MAILHIOT1992,Sun2013-N2},
O$_2$\cite{Sun2012}, CO\cite{Lipp1998,Bernard1998}, and
CO$_2$\cite{Iota1999,Iota2007}.

The triple bond in CO can be broken quite readily under pressure, and
it can polymerize at rather low pressures and temperatures.  For
example, Raman spectroscopy studies have found that molecular CO
polymerizes at a pressure of 4--5 GPa and temperatures $>$80 K
\cite{Cromer1983,Katz1984}.  The product reacts photochemically with
visible laser light and the transformation is reversible at ambient
conditions.  Polymerization of CO via the breaking of triple bonds
leading to the formation of C=C bonds has also been studied
\cite{Lipp1998, Evans2006}.  Lipp \textit{et al.}\ recently reported
that the Fourier transform infra-red (FTIR) spectrum of solid
polymeric CO (p-CO), which decomposes explosively into CO$_2$ and
glassy carbon, might be explained by rings containing
--C--O--(C=O)--C-- units.
Ceppatelli \textit{et al.}\ \cite{Ceppatelli2009} found that an
extended amorphous material forms from polycarbonyl chains at
temperatures $<$300 K, while above room temperature polycarbonyl
chains decompose into carbon dioxide and epoxy rings.

On the other hand, in previous work, a metallic zig-zag polymeric
chain-like CO material was found to be more stable than molecular CO
at ambient pressure \cite{Sun2011}.  This structure of $P2_1/m$
symmetry is formed from polycarbonyl chains containing a mixture of
single and double bonds.  
Although not crystalline, chain-like structures
consisting of five CO molecules have been successfully synthesized
using organic chemistry methods.  \cite{Rubin2000} 
However, according to Peierls' distortion theorem \cite{Peierls1979},
one-dimensional metallic chains are unstable to a distortion which
opens up a band gap between the occupied and unoccupied electronic
states.  Although weak interactions between chains in solid $P2_1/m$
persist, one might suspect that a Peierls' distortion to an insulating
structure could be energetically favorable in polymeric CO.

It is widely accepted that molecular gases such as CO, N$_2$ and
CO$_2$ are likely to be more stable at ambient pressure than their
polyermized counterparts. However, a very few exceptions have recently
been discovered.  For example, Wen\ \textit{et al.}\ \cite{Wen2011}\
found graphane sheets to be energetically more stable than benzene
under pressure.
Within the same 1:1 stoichiometry as CO, the crystal
structure of silicon monoxide was explored theoretically and similar
crystalline structures to those of CO were predicted. \cite{Alkaabi2014}
However, recent experiments on amorphous SiO found evidence of atomic-scale
disproportionation and suboxide-type tetrahedral
coordination. \cite{Hirata2016} 
It is also worth mentioning that CO is likely to be the second most
abundant molecule in interstellar space.\cite{Allamandola1999,Collings2003}
It is suggested to be present in dust or grains found in dense molecular
clouds, and low pressure polymeric structures of CO may therefore exist.

Using the \textit{ab initio} random structure searching (AIRSS) method
\cite{Pickard2006,Pickard2011}, we predict that several new CO
structures are energetically more stable than previously-known ones
over a wide range of pressures.  Among them we find a chain-like
polymeric phase consisting of six-membered rings (space group:
$Pna2_1$) at ambient pressures to be more stable than the molecular
phases.
More importantly, we find that polymeric $Pna2_1$-CO is a potential
``high energy density'' material, which can release about $4$--$9$
times more energy than Trinitrotoluene (TNT), if it decomposes into
carbon+CO$_2$ or reacts with oxygen and converts into CO$_2$.

\begin{figure}[tp]
\begin{center}
\includegraphics[width=0.5\textwidth]{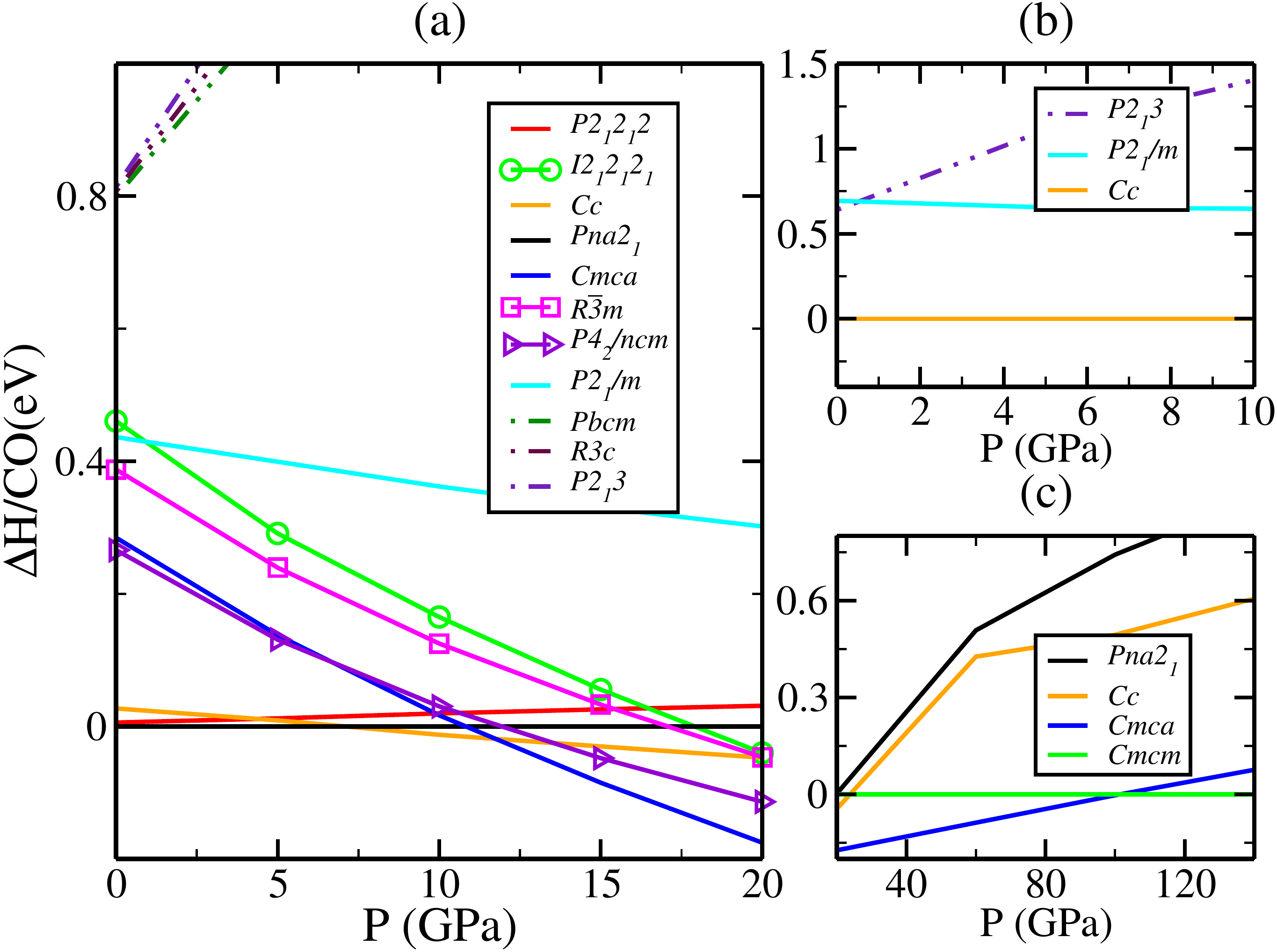}
\caption{Enthalpy-pressure relations for CO structures.  (a)
  Enthalpies of the most relevant structures at low pressures $<$20
  GPa relative to $Pna2_1$.  (b) Three representative structures (one
  molecular phase, two chain-like structures) at low pressures $<$10
  GPa relative to the $Cc$ phase calculated using the hybrid HSE06
  functional \cite{Heyd2004,Heyd2006,Paier2005,Paier2006}. The $Cc$
  structure is chosen as the reference due to the prohibitive cost of
  calculations for $Pna2_1$.  
(c) At high pressures up to 140 GPa relative to $Cmcm$. 
The kinks result from sudden volume changes during compression.
  Calculations for (a) and (c) were performed with the PBE functional
  together with the DFT-D3 correction using the Becke-Jonson (BJ)
  damping function parameters of Grimme \textit{et al.}
  \cite{Grimme2010,Grimme2011} Dashed lines represent the molecular
  phases, solid lines with symbols represent the framework and layered
  structures, and solid lines represent the chain-like structures.
}
\label{fig:enthalpy}
\end{center}
\end{figure}

Structure searches were performed using the CASTEP
\cite{CASTEP} code and the AIRSS approach, and structural
optimizations with higher accurcy criteria were performed with VASP
\cite{VASP} for cross checking.
Previous work has shown that dispersion corrections for molecular and
layered systems must be included to obtain accurate lattice constants
\cite{Lu2015,Zhou2016}.  We have investigated several commonly used
dispersion corrections implemented in the VASP \cite{VASP}, as listed in
TABLE\ \ref{table:potential}. The errors in the lattice constants with
the standard PBE functional are large, but they are much reduced when
a van der Waals corrected functional is used. Grimme's DFT-D3
(BJ-damping) method \cite{Grimme2010,Grimme2011} combined with the PBE
functional (PBE-D3) gives the smallest error for molecular systems
such as CO and CO$_2$, while the optB88 functional
\cite{Roman-Perez2009,Klimes2011,Thonhauser2007} together with the
vdW-DF corrections of Langreth and Lundqvist \textit{et al.}\
\cite{Dion2004} gives the most accurate results for the inter-layer
separations of graphite and crystalline MoS$_2$.

\begin{table*}[hpt]
\centering
 \begin{tabular}[t]{c|c|c|c|c|c}
    \hline
    \multicolumn {1}{c|}{exchange-correlation} &\multicolumn {1}{c|}{PBE/err.({\%})} &\multicolumn {1}{c|}{vdW-DF2/err.({\%})} &\multicolumn {1}{c|}{optB88-vdW/err.({\%})} &\multicolumn {1}{c|}{PBE-D3/err.({\%})} &\multicolumn {1}{c}{Exp.} \\
    \hline
    $\alpha$-CO & 6.093/8.2 & 5.502/2.3 & 5.447/3.3 & 5.634/0.07 & 5.630 \cite{Vegard1930}\ \\
    $\alpha$-CO$_2$ & 5.944/5.7 & 5.538/1.5 & 5.432/3.4 & 5.644/0.4 & 5.624 \cite{SIMON1980} \\
    graphite & 37.592/12.4 & 34.775/4.0 & 33.442/0.02 & 33.717/0.8 & 33.450 \cite{NIXON1966} \\
    MoS$_2$ & 13.399/9.0 & 12.956/5.4 & 12.500/1.7 & 12.077/1.8 & 12.294 \cite{BRONSEMA1986} \\
    \hline
  \end{tabular} 
  \caption{
    Lattice constants in \AA\ for a variety of solids calculated using VASP
    and different exchange-correlation functionals. The structures were optimised using
    DFT-D3(BJ-damping) \cite{Grimme2010,Grimme2011} and the PBE functional, and two
    other vdW-DF corrections (optB88-vdW \cite{Roman-Perez2009,Klimes2011,Thonhauser2007}
    and the rPW86 functional with the vdW-DF2 correction \cite{Dion2004,Lee2010})
    and the PBE functional \cite{Perdew1996}.
    Experimental lattice constants for the molecular $\alpha$-CO and $\alpha$-CO$_2$ phases
    were obtained from Vegard \textit{et al.} \cite{Vegard1930} and Simon \textit{et al.} \cite{SIMON1980}
    including interlayer distances of graphite from Nixon \textit{et al.} \cite{NIXON1966}
    and layered MoS$_2$ (space group $P6_3/mmc$) from Bronsema \textit{et al.} \cite{BRONSEMA1986}.
    For comparison, we list the absolute values and relative errors (err.) for each structure.
    For the molecular systems, the error arising from the DFT-D3 correction is
    smaller than that from the PBE functional and vdW-DF2 correction.
  }
\label{table:potential}
\end{table*}

  CO is relatively unstable compared to graphite and
  CO$_2$ in an oxygen rich environment. However, here we restrict
  ourselves to an oxygen poor environment and focus only on the 1:1
  stoichiometry.
Enthalpy-pressure relations for the most relevant CO structures are
shown in Fig.\ \ref{fig:enthalpy}.  Calculations using the PBE-D3 (BJ)
functional suggest that at least four structures, including
$P2_12_12$, $Pna2_1$, $Cc$, and $Cmca$ are more stable than the
previously predicted zig-zag chain-like $P2_1/m$ structure
\cite{Sun2011}.
Among them, as shown in Fig.\ \ref{fig:enthalpy}(a), the polymeric
chain-like $Pna2_1$ structure has an enthalpy of about 0.436 eV/CO
lower than that of $P2_1/m$, and about 0.814 eV/CO lower than the
molecular $\alpha$-CO structure (space group: $P2_13$) at 0 GPa.
$Pna2_1$-CO is predicted to transform into the $Cc$ structure at
pressures of about 7.1 GPa.  With further compression, another
chain-like structure of $Cmca$ symmetry is predicted to be stable in a
wide pressure range of about 12--100 GPa, as can be seen in Fig.\
\ref{fig:enthalpy}(c).  Calculations with the HSE06 hybrid functional
\cite{Heyd2004,Heyd2006,Paier2005,Paier2006} support the conclusion
that the $Cc$ structure is more stable than $P2_1/m$ and the molecular
$\alpha$-CO structure at pressures $<$10 GPa.  To investigate the
robustness of our results we calculated the enthalpies with commonly
used functionals including PBE, PBE+vdW-DF2, and optB88-vdW.  In each
case we found the polymeric chain-like $Pna2_1$ structure to be the
most stable phase at low pressures. Additional phase transitions occur
at higher pressures.

  Since $Pna2_1$-CO is thermodynamically much more
  favorable than molecular CO, it has a much higher energy density than
  molecular CO.
  We calculate that one kilogram of $Pna2_1$-CO can release about 16.2
  megajoules of energy when it decomposes into graphite and
  $\alpha$-CO$_2$.  (The energy of 1 kg TNT is about 4.2 megajoule.)
  If $Pna2_1$-CO reacts with oxygen and converts completely into
  CO$_2$, it can release up to about 37.3 megajoules of energy.

\begin{figure}[ht]
\begin{center}
\includegraphics[width=8.2cm]{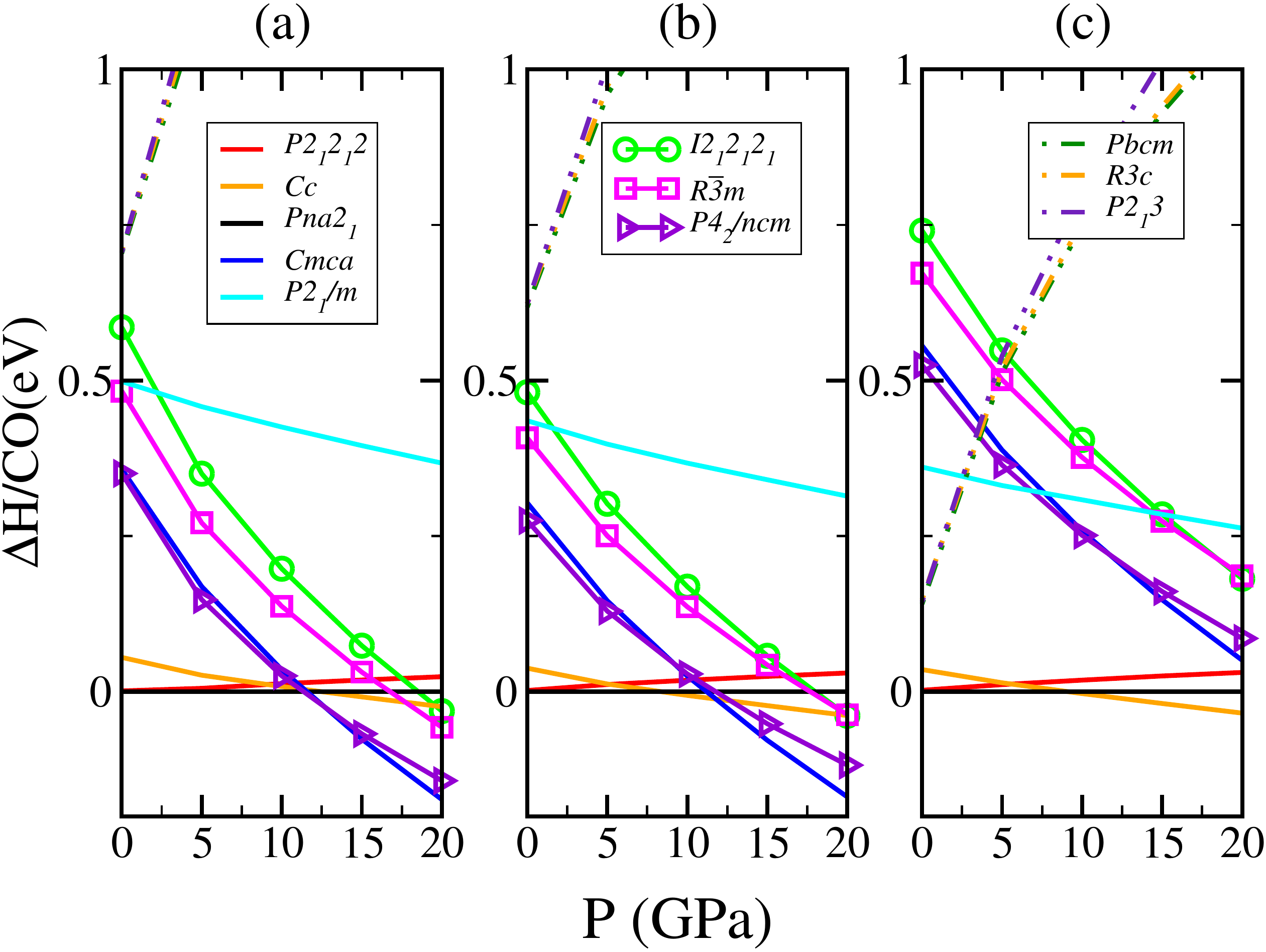}
\caption{
Static lattice enthalpy-pressure relations for CO structures calculated using various 
exchange-correlation functionals.
(a) PBE \cite{Perdew1996}, (b) optB88-vdW \cite{Roman-Perez2009,Klimes2011,Thonhauser2007}, 
(c) vdW-DF2 \cite{Dion2004,Lee2010}.  
Dashed lines represent the molecular phases, solid lines with marks represent the framework 
and layered structures, and solid lines represent the chain-like structures. Note that the 
labels are separated into three parts in the three panels.
}
\label{fig:enthalpy-l}
\end{center}
\end{figure}

\begin{figure}[hpt]
\begin{center}
\includegraphics[width=0.4\textwidth]{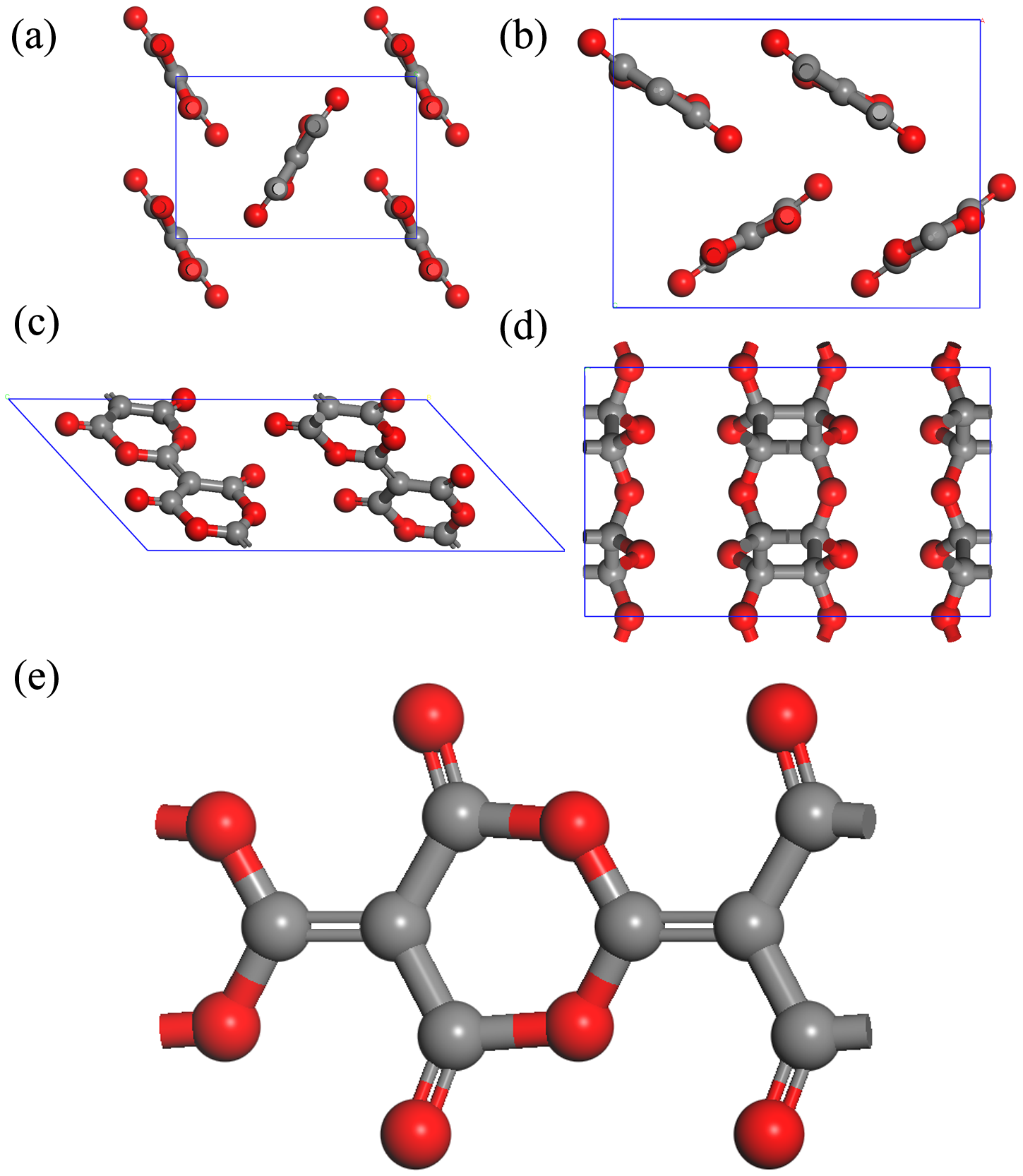}
\caption{
  The four best candidate structures for the ground state of polymeric
  CO at low pressures (grey color for C atom and red for O atom).  
  (a) $P2_12_12$ viewed from [010], (b) $Pna2_1$ from [001], $Cc$ from [010], (d) $Cmca$, 
(e) sketch of the single-chain with six-membered ring, 
which is the basic building block unit of $P2_12_12$, $Pna2_1$ and $Cc$.
}
\label{fig:CS}
\end{center}
\end{figure}

As shown in the crystal structure in Fig.\ \ref{fig:CS}(b), $Pna2_1$
is a chain-like structure consisting of six-membered rings connected
by C=C double bonds.  Each six-membered ring contains four carbon
atoms and two oxygen atoms, and two additional oxygen atoms are
attached to two carbon atoms in the ring to form two C=O carbonyl
groups,
sketched in Fig.\ \ref{fig:CS}(e).
Structures optimized at 0 GPa using the PBE-D3 (BJ)
functional have single-bond lengths in the range 1.309--1.455 \AA,
while the C=C bond length is 1.413 \AA\ and the shorter C=O bond is of
length 1.200 \AA.  As shown in Fig.\ \ref{fig:CS}(a) and (c), in
addition to the $Pna2_1$ phase we have found two other conformations
of similar structures with six-membered rings ($P2_12_12$ and $Cc$).
The higher pressure $Cmca$ structure is composed of units similar to
the single-bonded $I2_12_12_1$ phase, as can be seen in Fig.\
\ref{fig:CS}(d).  The optimized lattice parameters of the most
relevant structures are listed in Supplemental Material \cite{SM}
TABLE \uppercase\expandafter{\romannumeral1}.

The $Pna2_1$-CO structure (16 formula units (fu)), $P2_12_12$ (8 fu),
and $Cc$ (8 fu) have very similar structures featuring six-membered
rings. 
Phonon calculations for the $Pna2_1$ structure are expensive, and 
we have therefore instead calculated the phonon dispersion curves of the
closely related $P2_12_12$ structure at 0 GPa.  As shown in the
Supplemental Material \cite{SM} Fig.\ 2, $P2_12_12$ does not have any
negative (imaginary) phonon frequencies, and the structure is
predicted to be dynamically stable at ambient pressure.  The phonon
modes of the $P2_12_12$ phase can be divided into three groups.  The
high frequency bands at around 1394--1420 and 1756--1815 cm$^{-1}$
arise from C=C and C=O stretching within the chain,
which agrees with the FTIR-spectra measurements of Lipp 
\textit{et al.} \cite{Lipp2005}.
The intermediate frequency bands from 608 to 1207 cm$^{-1}$ arise from
the C-O and C-C single bonds in the plane of the six-membered rings,
while the frequencies below 602 cm$^{-1}$ mainly arise from
inter-chain vibrations. The dispersionless bands along the Z--T, Y--X
and U--R directions result from the parallel arrangement of the
chains, showing that the interactions between the chains are weak.
The dispersion relations are quite similar to those found in the
chain-like CO structures considered in earlier work \cite{Sun2011}.

\begin{figure}[hpt]
\begin{center}
\includegraphics[width=0.45\textwidth]{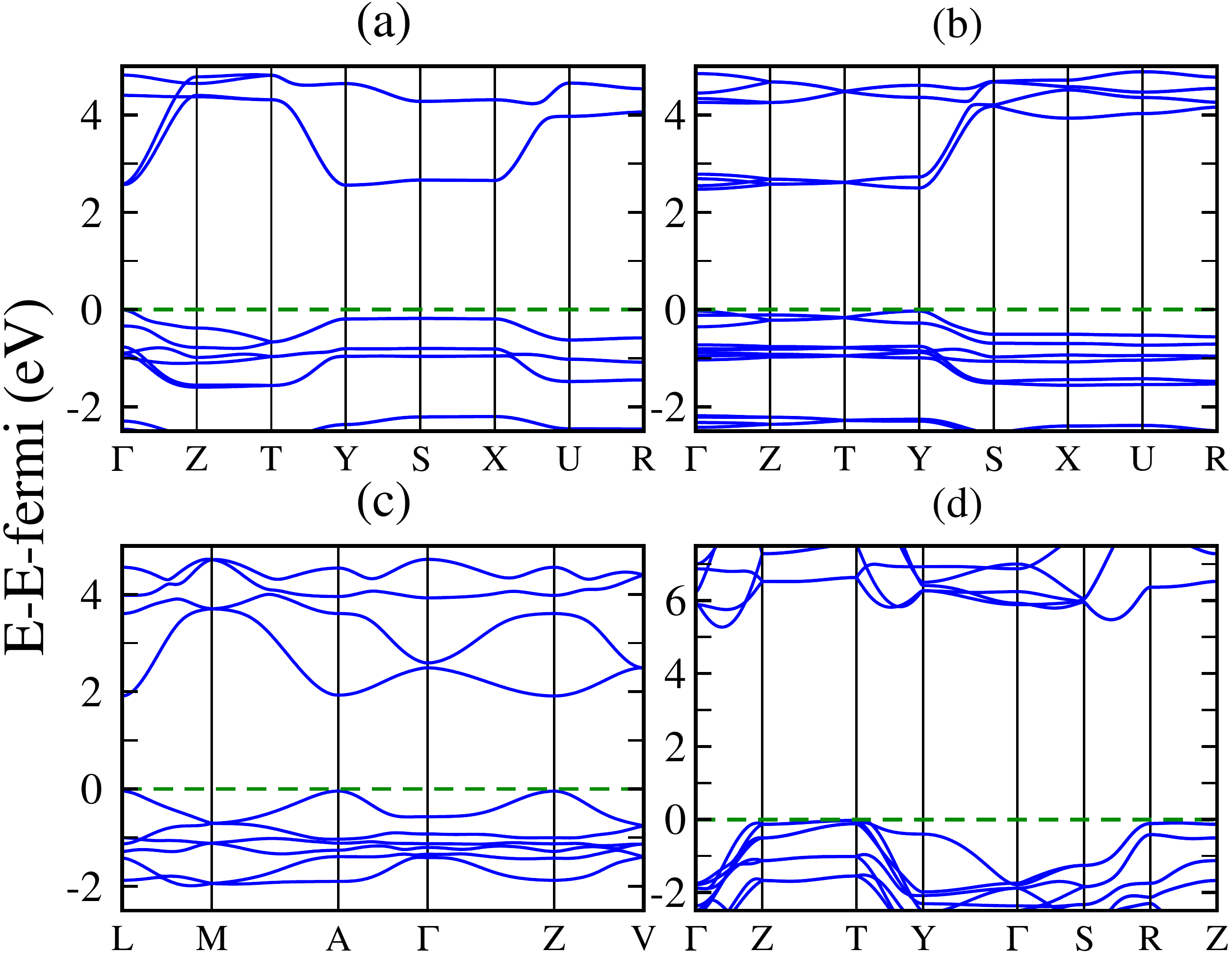}
\caption{
  Electronic band structures of four new polymeric phases, (a)
  $P2_12_12$ at 0 GPa, (b) $Pna2_1$ at 0 GPa, (c) $Cc$ at 10 GPa, and
  (d) $Cmca$ at 15 GPa.  The Fermi energy is shown as a dashed line at
  zero energy.  
These calculations were performed using 
the optimised PBE-D3 (BJ) functionals
  \cite{Grimme2010,Grimme2011}.  }
\label{fig:band}
\end{center}
\end{figure}

The electronic band structure and density of states of $P2_12_12$ and
$Pna2_1$ at 0 GPa, $Cc$ at 10 GPa and $Cmca$ at 15 GPa are shown in
Fig.\ \ref{fig:band}.  Calculations with the PBE-D3 (BJ) functional
\cite{Grimme2010,Grimme2011} suggest that they are all semiconductors
and that the polymeric chains $P2_12_12$ and $Pna2_1$ have band gaps
of about $2.5$--$2.6$ eV.  However, the hybrid HSE06 functional is
expected to give larger and more accurate gaps than semi-local
functionals, and we take the band gap of 3.6 eV obtained with the
HSE06 functional as our best estimate. We find that $P2_12_12$ and
$Pna2_1$ have direct band gaps at the $\Gamma$ point, and they may be
classified as large band-gap semiconductors and could be transparent.  
The bands of $P2_12_12$
along the Y--S--X direction at the top of the valence bands and bottom
of the conduction bands are almost dispersionless, which indicates
that the carriers have a large effective mass.  The large carrier
effective mass of $Pna2_1$ along a wide range of high symmetry
directions (except Y--S) results from the weak inter-chain
interactions that are similar to those in the $P2_12_12$ phase.

The enthalpy-pressure plots in Fig.\ \ref{fig:enthalpy} suggest that
the chain-like $Pna2_1$-CO is energetically stable compared to
molecular phases and the polycarbonyl chain-like phases predicted
previously at ambient pressure. The phonon spectra confirm that the
six-membered rings are stable.
It is well known that temperature can be an important factor in
determining the relative stability of structures \cite{Sun2011}.
We therefore tested the stability of the $Pna2_1$ phase at room
temperature using \textit{ab initio} molecular dynamics (AIMD)
simulations with Parrinello-Rahman ($NpT$) dynamics and a Langevin
thermostat \cite{PARRINELLO1980,PARRINELLO1981}.
As shown in the Supplemental Material \cite{SM} Fig.\ 4, the drift in
the statistical quantities during the 12 pico-second (ps) simulation
are small.
Due to the large cost of the AIMD calculations, we terminated the
simulation after 12 ps, which was considered to be sufficient to
confirm the stability of $Pna2_1$ at room temperature.

In summary, polymerization of molecular CO leads to the breaking of
triple CO bonds under increasing pressure, and subsequently to
structures with a mixture of double and single bonds, and at higher
pressures to the dominance of single bonds.  This leads to substantial
changes in the energy and density of the material with pressure, and
changes in the electronic structure.
  Although experimental polymerization of CO normally
  result in amorphous structures, careful control of the conditions
  may yield pure crystalline structures.

Extensive AIRSS calculations have led us to predict that several new
CO structures are energetically more stable than previously-known
structures over a wide range of pressures.  Among them we find a
chain-like polymeric phase consisting of six-membered rings (space
group: $Pna2_1$) to be the most stable at low pressures.  Using
results from several different functionals we conclude that the
$Pna2_1$ structure is more stable than molecular phases at ambient
pressure.  For example, results with the PBE-D3 (BJ) functional show
that the $Pna2_1$ structure is about 0.814 eV/CO lower in 
enthalpy than the molecular $\alpha$ phase. The
dynamical stability of the chain-like $Pna2_1$ phase is confirmed by
phonon calculations at 0 GPa and by an \textit{ab initio} molecular
dynamics simulation at ambient pressure and temperature. The $Pna2_1$
structure is an insulator at zero pressure with a large band gap that
we estimate to be about 3.6 eV.  $Pna2_1$-CO is a potential ``high
energy density'' material that can release about $4$--$9$ times as
much energy as TNT per mass.
\acknowledgments

J.S.\ is grateful for financial support from the MOST of China (Grant
Nos.\ 2016YFA0300404, 2015CB921202), 
the NSFC (Grant Nos.\ 51372112 and 11574133), the NSF of
Jiangsu Province (Grant No.\ BK20150012), Special Program for Applied
Research on Super Computation of the NSFC-Guangdong Joint Fund (the
second phase) and the Fundamental Research Funds for the Central
Universities.
C.J.P.\ and R.J.N.\ acknowledge financial support from the Engineering
and Physical Sciences Research Council (EPSRC) of U.K.\ under grants
[EP/G007489/2] (C.J.P.) and [EP/J017639/1] (R.J.N.).  C.J.P.\ is also
supported by the Royal Society through a Royal Society Wolfson
Research Merit award.
Some of the calculations were performed on the supercomputer 
in the High Performance Computing Center of Nanjing University.

\bibliographystyle{apsrev}
%\bibliography{hp-CO-5}

\begin{thebibliography}{53}
\expandafter\ifx\csname natexlab\endcsname\relax\def\natexlab#1{#1}\fi
\expandafter\ifx\csname bibnamefont\endcsname\relax
  \def\bibnamefont#1{#1}\fi
\expandafter\ifx\csname bibfnamefont\endcsname\relax
  \def\bibfnamefont#1{#1}\fi
\expandafter\ifx\csname citenamefont\endcsname\relax
  \def\citenamefont#1{#1}\fi
\expandafter\ifx\csname url\endcsname\relax
  \def\url#1{\texttt{#1}}\fi
\expandafter\ifx\csname urlprefix\endcsname\relax\def\urlprefix{URL }\fi
\providecommand{\bibinfo}[2]{#2}
\providecommand{\eprint}[2][]{\url{#2}}

\bibitem[{\citenamefont{Hemley}(2000)}]{Hemley2000}
\bibinfo{author}{\bibfnamefont{R.~J.} \bibnamefont{Hemley}},
  \bibinfo{journal}{Annu.~Rev.~Phys.~Chem} \textbf{\bibinfo{volume}{51}},
  \bibinfo{pages}{763} (\bibinfo{year}{2000}).

\bibitem[{\citenamefont{Schettino and Bini}(2003)}]{Schettino2003}
\bibinfo{author}{\bibfnamefont{V.}~\bibnamefont{Schettino}} \bibnamefont{and}
  \bibinfo{author}{\bibfnamefont{R.}~\bibnamefont{Bini}},
  \bibinfo{journal}{Phys. Chem. Chem. Phys.} \textbf{\bibinfo{volume}{5}},
  \bibinfo{pages}{1951} (\bibinfo{year}{2003}).

\bibitem[{\citenamefont{Lipp et~al.}(2005)\citenamefont{Lipp, Evans, Baer, and
  Yoo}}]{Lipp2005}
\bibinfo{author}{\bibfnamefont{M.~J.} \bibnamefont{Lipp}},
  \bibinfo{author}{\bibfnamefont{W.~J.} \bibnamefont{Evans}},
  \bibinfo{author}{\bibfnamefont{B.~J.} \bibnamefont{Baer}}, \bibnamefont{and}
  \bibinfo{author}{\bibfnamefont{C.~S.} \bibnamefont{Yoo}},
  \bibinfo{journal}{Nature~Mat.} \textbf{\bibinfo{volume}{4}},
  \bibinfo{pages}{211} (\bibinfo{year}{2005}).

\bibitem[{\citenamefont{Yoo et~al.}(1999)\citenamefont{Yoo, Cynn, Gygi, Galli,
  Iota, Nicol, Carlson, Hausermann, and Mailhiot}}]{Yoo1999}
\bibinfo{author}{\bibfnamefont{C.~S.} \bibnamefont{Yoo}},
  \bibinfo{author}{\bibfnamefont{H.}~\bibnamefont{Cynn}},
  \bibinfo{author}{\bibfnamefont{F.}~\bibnamefont{Gygi}},
  \bibinfo{author}{\bibfnamefont{G.}~\bibnamefont{Galli}},
  \bibinfo{author}{\bibfnamefont{V.}~\bibnamefont{Iota}},
  \bibinfo{author}{\bibfnamefont{M.}~\bibnamefont{Nicol}},
  \bibinfo{author}{\bibfnamefont{S.}~\bibnamefont{Carlson}},
  \bibinfo{author}{\bibfnamefont{D.}~\bibnamefont{Hausermann}},
  \bibnamefont{and} \bibinfo{author}{\bibfnamefont{C.}~\bibnamefont{Mailhiot}},
  \bibinfo{journal}{Phys.~Rev.~Lett.} \textbf{\bibinfo{volume}{83}},
  \bibinfo{pages}{5527} (\bibinfo{year}{1999}).

\bibitem[{\citenamefont{Eremets et~al.}(2004)\citenamefont{Eremets, Gavriliuk,
  Serebryanaya, Trojan, Dzivenko, Boehler, Mao, and Hemley}}]{Eremets2004}
\bibinfo{author}{\bibfnamefont{M.~I.} \bibnamefont{Eremets}},
  \bibinfo{author}{\bibfnamefont{A.~G.} \bibnamefont{Gavriliuk}},
  \bibinfo{author}{\bibfnamefont{N.~R.} \bibnamefont{Serebryanaya}},
  \bibinfo{author}{\bibfnamefont{I.~A.} \bibnamefont{Trojan}},
  \bibinfo{author}{\bibfnamefont{D.~A.} \bibnamefont{Dzivenko}},
  \bibinfo{author}{\bibfnamefont{R.}~\bibnamefont{Boehler}},
  \bibinfo{author}{\bibfnamefont{H.~K.} \bibnamefont{Mao}}, \bibnamefont{and}
  \bibinfo{author}{\bibfnamefont{R.~J.} \bibnamefont{Hemley}},
  \bibinfo{journal}{J.~Chem.~Phys.} \textbf{\bibinfo{volume}{121}},
  \bibinfo{pages}{11296} (\bibinfo{year}{2004}).

\bibitem[{\citenamefont{Cromer et~al.}(1983)\citenamefont{Cromer, Schiferl,
  Lesar, and Mills}}]{Cromer1983}
\bibinfo{author}{\bibfnamefont{D.~T.} \bibnamefont{Cromer}},
  \bibinfo{author}{\bibfnamefont{D.}~\bibnamefont{Schiferl}},
  \bibinfo{author}{\bibfnamefont{R.}~\bibnamefont{Lesar}}, \bibnamefont{and}
  \bibinfo{author}{\bibfnamefont{R.~L.} \bibnamefont{Mills}},
  \bibinfo{journal}{Acta Cryst. Section C} \textbf{\bibinfo{volume}{39}},
  \bibinfo{pages}{1146} (\bibinfo{year}{1983}).

\bibitem[{\citenamefont{Mills et~al.}(1984)\citenamefont{Mills, Schiferl, Katz,
  and Olinger}}]{Mills1984}
\bibinfo{author}{\bibfnamefont{R.~L.} \bibnamefont{Mills}},
  \bibinfo{author}{\bibfnamefont{D.}~\bibnamefont{Schiferl}},
  \bibinfo{author}{\bibfnamefont{A.~I.} \bibnamefont{Katz}}, \bibnamefont{and}
  \bibinfo{author}{\bibfnamefont{B.~W.} \bibnamefont{Olinger}},
  \bibinfo{journal}{Journal de Physique} \textbf{\bibinfo{volume}{45}},
  \bibinfo{pages}{187} (\bibinfo{year}{1984}).

\bibitem[{\citenamefont{Katz et~al.}(1984)\citenamefont{Katz, Schiferl, and
  Mills}}]{Katz1984}
\bibinfo{author}{\bibfnamefont{A.~I.} \bibnamefont{Katz}},
  \bibinfo{author}{\bibfnamefont{D.}~\bibnamefont{Schiferl}}, \bibnamefont{and}
  \bibinfo{author}{\bibfnamefont{R.~L.} \bibnamefont{Mills}},
  \bibinfo{journal}{J.~Phys.~Chem.} \textbf{\bibinfo{volume}{88}},
  \bibinfo{pages}{3176} (\bibinfo{year}{1984}).

\bibitem[{\citenamefont{Mills et~al.}(1986)\citenamefont{Mills, Olinger, and
  Cromer}}]{Mills1986}
\bibinfo{author}{\bibfnamefont{R.~L.} \bibnamefont{Mills}},
  \bibinfo{author}{\bibfnamefont{B.}~\bibnamefont{Olinger}}, \bibnamefont{and}
  \bibinfo{author}{\bibfnamefont{D.~T.} \bibnamefont{Cromer}},
  \bibinfo{journal}{J.~Chem.~Phys.} \textbf{\bibinfo{volume}{84}},
  \bibinfo{pages}{2837} (\bibinfo{year}{1986}).

\bibitem[{\citenamefont{Fracassi et~al.}(1986)\citenamefont{Fracassi, Cardini,
  Oshea, Impey, and Klein}}]{Fracassi1986}
\bibinfo{author}{\bibfnamefont{P.~F.} \bibnamefont{Fracassi}},
  \bibinfo{author}{\bibfnamefont{G.}~\bibnamefont{Cardini}},
  \bibinfo{author}{\bibfnamefont{S.}~\bibnamefont{Oshea}},
  \bibinfo{author}{\bibfnamefont{R.~W.} \bibnamefont{Impey}}, \bibnamefont{and}
  \bibinfo{author}{\bibfnamefont{M.~L.} \bibnamefont{Klein}},
  \bibinfo{journal}{Phys.~Rev.~B} \textbf{\bibinfo{volume}{33}},
  \bibinfo{pages}{3441} (\bibinfo{year}{1986}).

\bibitem[{\citenamefont{Frapper et~al.}(1997)\citenamefont{Frapper, Cu,
  Kertesz, Halet, Saillard, and Kertesz}}]{Frapper1997}
\bibinfo{author}{\bibfnamefont{G.}~\bibnamefont{Frapper}},
  \bibinfo{author}{\bibfnamefont{C.-X.} \bibnamefont{Cu}},
  \bibinfo{author}{\bibfnamefont{M.}~\bibnamefont{Kertesz}},
  \bibinfo{author}{\bibfnamefont{J.-F.} \bibnamefont{Halet}},
  \bibinfo{author}{\bibfnamefont{J.-Y.} \bibnamefont{Saillard}},
  \bibnamefont{and} \bibinfo{author}{\bibfnamefont{M.}~\bibnamefont{Kertesz}},
  \bibinfo{journal}{Chem.~Commun.} pp. \bibinfo{pages}{2011--2012}
  (\bibinfo{year}{1997}).

\bibitem[{\citenamefont{Lipp et~al.}(1998)\citenamefont{Lipp, Evans,
  Garcia-Baonza, and Lorenzana}}]{Lipp1998}
\bibinfo{author}{\bibfnamefont{M.}~\bibnamefont{Lipp}},
  \bibinfo{author}{\bibfnamefont{W.~J.} \bibnamefont{Evans}},
  \bibinfo{author}{\bibfnamefont{V.}~\bibnamefont{Garcia-Baonza}},
  \bibnamefont{and} \bibinfo{author}{\bibfnamefont{H.~E.}
  \bibnamefont{Lorenzana}}, \bibinfo{journal}{J. Low Temp. Phys.}
  \textbf{\bibinfo{volume}{111}}, \bibinfo{pages}{247} (\bibinfo{year}{1998}).

\bibitem[{\citenamefont{Bernard et~al.}(1998)\citenamefont{Bernard, Chiarotti,
  Scandolo, and Tosatti}}]{Bernard1998}
\bibinfo{author}{\bibfnamefont{S.}~\bibnamefont{Bernard}},
  \bibinfo{author}{\bibfnamefont{G.~L.} \bibnamefont{Chiarotti}},
  \bibinfo{author}{\bibfnamefont{S.}~\bibnamefont{Scandolo}}, \bibnamefont{and}
  \bibinfo{author}{\bibfnamefont{E.}~\bibnamefont{Tosatti}},
  \bibinfo{journal}{Phys.~Rev.~Lett.} \textbf{\bibinfo{volume}{81}},
  \bibinfo{pages}{2092} (\bibinfo{year}{1998}).

\bibitem[{\citenamefont{Evans et~al.}(2006)\citenamefont{Evans, Lipp, Yoo,
  Cynn, Herberg, Maxwell, and Nicol}}]{Evans2006}
\bibinfo{author}{\bibfnamefont{W.~J.} \bibnamefont{Evans}},
  \bibinfo{author}{\bibfnamefont{M.~J.} \bibnamefont{Lipp}},
  \bibinfo{author}{\bibfnamefont{C.~S.} \bibnamefont{Yoo}},
  \bibinfo{author}{\bibfnamefont{H.}~\bibnamefont{Cynn}},
  \bibinfo{author}{\bibfnamefont{J.~L.} \bibnamefont{Herberg}},
  \bibinfo{author}{\bibfnamefont{R.~S.} \bibnamefont{Maxwell}},
  \bibnamefont{and} \bibinfo{author}{\bibfnamefont{M.~F.} \bibnamefont{Nicol}},
  \bibinfo{journal}{Chem.~Mat.} \textbf{\bibinfo{volume}{18}},
  \bibinfo{pages}{2520} (\bibinfo{year}{2006}).

\bibitem[{\citenamefont{Ceppatelli et~al.}(2009)\citenamefont{Ceppatelli, Bini,
  and Schettino}}]{Ceppatelli2009}
\bibinfo{author}{\bibfnamefont{M.}~\bibnamefont{Ceppatelli}},
  \bibinfo{author}{\bibfnamefont{R.}~\bibnamefont{Bini}}, \bibnamefont{and}
  \bibinfo{author}{\bibfnamefont{V.}~\bibnamefont{Schettino}},
  \bibinfo{journal}{J.~Phys.~Chem. B} \textbf{\bibinfo{volume}{113}},
  \bibinfo{pages}{14640} (\bibinfo{year}{2009}).

\bibitem[{\citenamefont{Sun et~al.}(2011)\citenamefont{Sun, Klug, Pickard, and
  Needs}}]{Sun2011}
\bibinfo{author}{\bibfnamefont{J.}~\bibnamefont{Sun}},
  \bibinfo{author}{\bibfnamefont{D.~D.} \bibnamefont{Klug}},
  \bibinfo{author}{\bibfnamefont{C.~J.} \bibnamefont{Pickard}},
  \bibnamefont{and} \bibinfo{author}{\bibfnamefont{R.~J.} \bibnamefont{Needs}},
  \bibinfo{journal}{Phys.~Rev.~Lett.} \textbf{\bibinfo{volume}{106}},
  \bibinfo{pages}{145502} (\bibinfo{year}{2011}).

\bibitem[{\citenamefont{Mailhiot et~al.}(1992)\citenamefont{Mailhiot, Yang, and
  Mcmahan}}]{MAILHIOT1992}
\bibinfo{author}{\bibfnamefont{C.}~\bibnamefont{Mailhiot}},
  \bibinfo{author}{\bibfnamefont{L.~H.} \bibnamefont{Yang}}, \bibnamefont{and}
  \bibinfo{author}{\bibfnamefont{A.~K.} \bibnamefont{Mcmahan}},
  \bibinfo{journal}{Phys.~Rev.~B} \textbf{\bibinfo{volume}{46}},
  \bibinfo{pages}{14419} (\bibinfo{year}{1992}).

\bibitem[{\citenamefont{Sun et~al.}(2013)\citenamefont{Sun, Martinez-Canales,
  Klug, Pickard, and Needs}}]{Sun2013-N2}
\bibinfo{author}{\bibfnamefont{J.}~\bibnamefont{Sun}},
  \bibinfo{author}{\bibfnamefont{M.}~\bibnamefont{Martinez-Canales}},
  \bibinfo{author}{\bibfnamefont{D.~D.} \bibnamefont{Klug}},
  \bibinfo{author}{\bibfnamefont{C.~J.} \bibnamefont{Pickard}},
  \bibnamefont{and} \bibinfo{author}{\bibfnamefont{R.~J.} \bibnamefont{Needs}},
  \bibinfo{journal}{Phys. Rev. Lett.} \textbf{\bibinfo{volume}{111}},
  \bibinfo{pages}{175502} (\bibinfo{year}{2013}).

\bibitem[{\citenamefont{Sun et~al.}(2012)\citenamefont{Sun, Martinez-Canales,
  Klug, Pickard, and Needs}}]{Sun2012}
\bibinfo{author}{\bibfnamefont{J.}~\bibnamefont{Sun}},
  \bibinfo{author}{\bibfnamefont{M.}~\bibnamefont{Martinez-Canales}},
  \bibinfo{author}{\bibfnamefont{D.~D.} \bibnamefont{Klug}},
  \bibinfo{author}{\bibfnamefont{C.~J.} \bibnamefont{Pickard}},
  \bibnamefont{and} \bibinfo{author}{\bibfnamefont{R.~J.} \bibnamefont{Needs}},
  \bibinfo{journal}{Phys.~Rev.~Lett.} \textbf{\bibinfo{volume}{108}},
  \bibinfo{pages}{045503} (\bibinfo{year}{2012}).

\bibitem[{\citenamefont{Iota et~al.}(1999)\citenamefont{Iota, Yoo, and
  Cynn}}]{Iota1999}
\bibinfo{author}{\bibfnamefont{V.}~\bibnamefont{Iota}},
  \bibinfo{author}{\bibfnamefont{C.~S.} \bibnamefont{Yoo}}, \bibnamefont{and}
  \bibinfo{author}{\bibfnamefont{H.}~\bibnamefont{Cynn}},
  \bibinfo{journal}{Science} \textbf{\bibinfo{volume}{283}},
  \bibinfo{pages}{1510} (\bibinfo{year}{1999}).

\bibitem[{\citenamefont{Iota et~al.}(2007)\citenamefont{Iota, Yoo, Klepeis,
  Jenei, Evans, and Cynn}}]{Iota2007}
\bibinfo{author}{\bibfnamefont{V.}~\bibnamefont{Iota}},
  \bibinfo{author}{\bibfnamefont{C.~S.} \bibnamefont{Yoo}},
  \bibinfo{author}{\bibfnamefont{J.~H.} \bibnamefont{Klepeis}},
  \bibinfo{author}{\bibfnamefont{Z.}~\bibnamefont{Jenei}},
  \bibinfo{author}{\bibfnamefont{W.}~\bibnamefont{Evans}}, \bibnamefont{and}
  \bibinfo{author}{\bibfnamefont{H.}~\bibnamefont{Cynn}},
  \bibinfo{journal}{Nature~Mat.} \textbf{\bibinfo{volume}{6}},
  \bibinfo{pages}{34} (\bibinfo{year}{2007}).

\bibitem[{\citenamefont{Rubin and Gleiter}(2000)}]{Rubin2000}
\bibinfo{author}{\bibfnamefont{M.~B.} \bibnamefont{Rubin}} \bibnamefont{and}
  \bibinfo{author}{\bibfnamefont{R.}~\bibnamefont{Gleiter}},
  \bibinfo{journal}{Chem. Rev.} \textbf{\bibinfo{volume}{100}},
  \bibinfo{pages}{1121} (\bibinfo{year}{2000}).

\bibitem[{\citenamefont{Peierls}(1979)}]{Peierls1979}
\bibinfo{author}{\bibfnamefont{R.}~\bibnamefont{Peierls}},
  \emph{\bibinfo{title}{Surprises in theoretical physics}}
  (\bibinfo{publisher}{Princeton, N.J.: Princeton University Press},
  \bibinfo{year}{1979}).

\bibitem[{\citenamefont{Wen et~al.}(2011)\citenamefont{Wen, Hand, Labet, Yang,
  Hoffmann, Ashcroft, Oganov, and Lyakhov}}]{Wen2011}
\bibinfo{author}{\bibfnamefont{X.-D.} \bibnamefont{Wen}},
  \bibinfo{author}{\bibfnamefont{L.}~\bibnamefont{Hand}},
  \bibinfo{author}{\bibfnamefont{V.}~\bibnamefont{Labet}},
  \bibinfo{author}{\bibfnamefont{T.}~\bibnamefont{Yang}},
  \bibinfo{author}{\bibfnamefont{R.}~\bibnamefont{Hoffmann}},
  \bibinfo{author}{\bibfnamefont{N.~W.} \bibnamefont{Ashcroft}},
  \bibinfo{author}{\bibfnamefont{A.~R.} \bibnamefont{Oganov}},
  \bibnamefont{and} \bibinfo{author}{\bibfnamefont{A.~O.}
  \bibnamefont{Lyakhov}}, \bibinfo{journal}{Proc. Natl. Acad. Sci. U.S.A.}
  \textbf{\bibinfo{volume}{108}}, \bibinfo{pages}{6833} (\bibinfo{year}{2011}).

\bibitem[{\citenamefont{AlKaabi et~al.}(2014)\citenamefont{AlKaabi, Prasad,
  Kroll, Ashcroft, and Hoffmann}}]{Alkaabi2014}
\bibinfo{author}{\bibfnamefont{K.}~\bibnamefont{AlKaabi}},
  \bibinfo{author}{\bibfnamefont{D.~L. V.~K.} \bibnamefont{Prasad}},
  \bibinfo{author}{\bibfnamefont{P.}~\bibnamefont{Kroll}},
  \bibinfo{author}{\bibfnamefont{N.~W.} \bibnamefont{Ashcroft}},
  \bibnamefont{and} \bibinfo{author}{\bibfnamefont{R.}~\bibnamefont{Hoffmann}},
  \bibinfo{journal}{J.~Am.~Chem.~Soc.} \textbf{\bibinfo{volume}{136}},
  \bibinfo{pages}{3410} (\bibinfo{year}{2014}).

\bibitem[{\citenamefont{Hirata et~al.}(2016)\citenamefont{Hirata, Kohara,
  Asada, Arao, Yogi, Imai, Tan, Fujita, and Chen}}]{Hirata2016}
\bibinfo{author}{\bibfnamefont{A.}~\bibnamefont{Hirata}},
  \bibinfo{author}{\bibfnamefont{S.}~\bibnamefont{Kohara}},
  \bibinfo{author}{\bibfnamefont{T.}~\bibnamefont{Asada}},
  \bibinfo{author}{\bibfnamefont{M.}~\bibnamefont{Arao}},
  \bibinfo{author}{\bibfnamefont{C.}~\bibnamefont{Yogi}},
  \bibinfo{author}{\bibfnamefont{H.}~\bibnamefont{Imai}},
  \bibinfo{author}{\bibfnamefont{Y.}~\bibnamefont{Tan}},
  \bibinfo{author}{\bibfnamefont{T.}~\bibnamefont{Fujita}}, \bibnamefont{and}
  \bibinfo{author}{\bibfnamefont{M.}~\bibnamefont{Chen}},
  \bibinfo{journal}{Nat. Commun.} \textbf{\bibinfo{volume}{7}},
  \bibinfo{pages}{11591} (\bibinfo{year}{2016}).

\bibitem[{\citenamefont{Allamandola et~al.}(1999)\citenamefont{Allamandola,
  Bernstein, Sandford, and Walker}}]{Allamandola1999}
\bibinfo{author}{\bibfnamefont{L.}~\bibnamefont{Allamandola}},
  \bibinfo{author}{\bibfnamefont{M.}~\bibnamefont{Bernstein}},
  \bibinfo{author}{\bibfnamefont{S.}~\bibnamefont{Sandford}}, \bibnamefont{and}
  \bibinfo{author}{\bibfnamefont{R.}~\bibnamefont{Walker}},
  \bibinfo{journal}{Space Sci. Rev.} \textbf{\bibinfo{volume}{90}},
  \bibinfo{pages}{219} (\bibinfo{year}{1999}), ISSN \bibinfo{issn}{0038-6308},
  \bibinfo{note}{iSSI Workshop on Composition and Origin of Cometary Material,
  BERN, SWITZERLAND, SEP 14-18, 1998}.

\bibitem[{\citenamefont{Collings et~al.}(2003)\citenamefont{Collings, Dever,
  Fraser, McCoustra, and Williams}}]{Collings2003}
\bibinfo{author}{\bibfnamefont{M.}~\bibnamefont{Collings}},
  \bibinfo{author}{\bibfnamefont{J.}~\bibnamefont{Dever}},
  \bibinfo{author}{\bibfnamefont{H.}~\bibnamefont{Fraser}},
  \bibinfo{author}{\bibfnamefont{M.}~\bibnamefont{McCoustra}},
  \bibnamefont{and} \bibinfo{author}{\bibfnamefont{D.}~\bibnamefont{Williams}},
  \bibinfo{journal}{Astrophys. J.} \textbf{\bibinfo{volume}{583}},
  \bibinfo{pages}{1058} (\bibinfo{year}{2003}), ISSN \bibinfo{issn}{0004-637X}.

\bibitem[{\citenamefont{Pickard and Needs}(2006)}]{Pickard2006}
\bibinfo{author}{\bibfnamefont{C.~J.} \bibnamefont{Pickard}} \bibnamefont{and}
  \bibinfo{author}{\bibfnamefont{R.~J.} \bibnamefont{Needs}},
  \bibinfo{journal}{Phys.~Rev.~Lett.} \textbf{\bibinfo{volume}{97}},
  \bibinfo{pages}{045504} (\bibinfo{year}{2006}).

\bibitem[{\citenamefont{Pickard and Needs}(2011)}]{Pickard2011}
\bibinfo{author}{\bibfnamefont{C.~J.} \bibnamefont{Pickard}} \bibnamefont{and}
  \bibinfo{author}{\bibfnamefont{R.~J.} \bibnamefont{Needs}},
  \bibinfo{journal}{J. Phys.: Conden. Matter} \textbf{\bibinfo{volume}{23}},
  \bibinfo{pages}{053201} (\bibinfo{year}{2011}).

\bibitem[{\citenamefont{Heyd and Scuseria}(2004)}]{Heyd2004}
\bibinfo{author}{\bibfnamefont{J.}~\bibnamefont{Heyd}} \bibnamefont{and}
  \bibinfo{author}{\bibfnamefont{G.~E.} \bibnamefont{Scuseria}},
  \bibinfo{journal}{J.~Chem.~Phys.} \textbf{\bibinfo{volume}{121}},
  \bibinfo{pages}{1187} (\bibinfo{year}{2004}).

\bibitem[{\citenamefont{Heyd et~al.}(2006)\citenamefont{Heyd, Scuseria, and
  Ernzerhof}}]{Heyd2006}
\bibinfo{author}{\bibfnamefont{J.}~\bibnamefont{Heyd}},
  \bibinfo{author}{\bibfnamefont{G.~E.} \bibnamefont{Scuseria}},
  \bibnamefont{and}
  \bibinfo{author}{\bibfnamefont{M.}~\bibnamefont{Ernzerhof}},
  \bibinfo{journal}{J.~Chem.~Phys.} \textbf{\bibinfo{volume}{124}},
  \bibinfo{pages}{219906} (\bibinfo{year}{2006}).

\bibitem[{\citenamefont{Paier et~al.}(2005)\citenamefont{Paier, Hirschl,
  Marsman, and Kresse}}]{Paier2005}
\bibinfo{author}{\bibfnamefont{J.}~\bibnamefont{Paier}},
  \bibinfo{author}{\bibfnamefont{R.}~\bibnamefont{Hirschl}},
  \bibinfo{author}{\bibfnamefont{M.}~\bibnamefont{Marsman}}, \bibnamefont{and}
  \bibinfo{author}{\bibfnamefont{G.}~\bibnamefont{Kresse}},
  \bibinfo{journal}{J.~Chem.~Phys.} \textbf{\bibinfo{volume}{122}},
  \bibinfo{pages}{234102} (\bibinfo{year}{2005}).

\bibitem[{\citenamefont{Paier et~al.}(2006)\citenamefont{Paier, Marsman,
  Hummer, Kresse, Gerber, and Angyan}}]{Paier2006}
\bibinfo{author}{\bibfnamefont{J.}~\bibnamefont{Paier}},
  \bibinfo{author}{\bibfnamefont{M.}~\bibnamefont{Marsman}},
  \bibinfo{author}{\bibfnamefont{K.}~\bibnamefont{Hummer}},
  \bibinfo{author}{\bibfnamefont{G.}~\bibnamefont{Kresse}},
  \bibinfo{author}{\bibfnamefont{I.~C.} \bibnamefont{Gerber}},
  \bibnamefont{and} \bibinfo{author}{\bibfnamefont{J.~G.}
  \bibnamefont{Angyan}}, \bibinfo{journal}{J.~Chem.~Phys.}
  \textbf{\bibinfo{volume}{124}}, \bibinfo{pages}{154709}
  (\bibinfo{year}{2006}).

\bibitem[{\citenamefont{Grimme et~al.}(2010)\citenamefont{Grimme, Antony,
  Ehrlich, and Krieg}}]{Grimme2010}
\bibinfo{author}{\bibfnamefont{S.}~\bibnamefont{Grimme}},
  \bibinfo{author}{\bibfnamefont{J.}~\bibnamefont{Antony}},
  \bibinfo{author}{\bibfnamefont{S.}~\bibnamefont{Ehrlich}}, \bibnamefont{and}
  \bibinfo{author}{\bibfnamefont{H.}~\bibnamefont{Krieg}},
  \bibinfo{journal}{J.~Chem.~Phys.} \textbf{\bibinfo{volume}{132}},
  \bibinfo{pages}{154104} (\bibinfo{year}{2010}).

\bibitem[{\citenamefont{Grimme et~al.}(2011)\citenamefont{Grimme, Ehrlich, and
  Goerigk}}]{Grimme2011}
\bibinfo{author}{\bibfnamefont{S.}~\bibnamefont{Grimme}},
  \bibinfo{author}{\bibfnamefont{S.}~\bibnamefont{Ehrlich}}, \bibnamefont{and}
  \bibinfo{author}{\bibfnamefont{L.}~\bibnamefont{Goerigk}},
  \bibinfo{journal}{J. Comput. Chem.} \textbf{\bibinfo{volume}{32}},
  \bibinfo{pages}{1456} (\bibinfo{year}{2011}).

\bibitem[{\citenamefont{Clark et~al.}(2005)\citenamefont{Clark, Segall,
  Pickard, Hasnip, Probert, Refson, and Payne}}]{CASTEP}
\bibinfo{author}{\bibfnamefont{S.~J.} \bibnamefont{Clark}},
  \bibinfo{author}{\bibfnamefont{M.~D.} \bibnamefont{Segall}},
  \bibinfo{author}{\bibfnamefont{C.~J.} \bibnamefont{Pickard}},
  \bibinfo{author}{\bibfnamefont{P.~J.} \bibnamefont{Hasnip}},
  \bibinfo{author}{\bibfnamefont{M.~J.} \bibnamefont{Probert}},
  \bibinfo{author}{\bibfnamefont{K.}~\bibnamefont{Refson}}, \bibnamefont{and}
  \bibinfo{author}{\bibfnamefont{M.~C.} \bibnamefont{Payne}},
  \bibinfo{journal}{Z. Kristallogr.} \textbf{\bibinfo{volume}{220}},
  \bibinfo{pages}{567} (\bibinfo{year}{2005}).

\bibitem[{\citenamefont{Kresse and Furthm\"{u}ller}(1996)}]{VASP}
\bibinfo{author}{\bibfnamefont{G.}~\bibnamefont{Kresse}} \bibnamefont{and}
  \bibinfo{author}{\bibfnamefont{J.}~\bibnamefont{Furthm\"{u}ller}},
  \bibinfo{journal}{Comp.~Mat.~Sci.} \textbf{\bibinfo{volume}{6}},
  \bibinfo{pages}{15} (\bibinfo{year}{1996}).

\bibitem[{\citenamefont{{Lu} et~al.}(2015)\citenamefont{{Lu}, {Kim}, {Yang},
  {Gao}, {Wu}, {Shao}, {Li}, {Zhou}, {Sun}, {Akinwande} et~al.}}]{Lu2015}
\bibinfo{author}{\bibfnamefont{P.}~\bibnamefont{{Lu}}},
  \bibinfo{author}{\bibfnamefont{J.-S.} \bibnamefont{{Kim}}},
  \bibinfo{author}{\bibfnamefont{J.}~\bibnamefont{{Yang}}},
  \bibinfo{author}{\bibfnamefont{H.}~\bibnamefont{{Gao}}},
  \bibinfo{author}{\bibfnamefont{J.}~\bibnamefont{{Wu}}},
  \bibinfo{author}{\bibfnamefont{D.}~\bibnamefont{{Shao}}},
  \bibinfo{author}{\bibfnamefont{B.}~\bibnamefont{{Li}}},
  \bibinfo{author}{\bibfnamefont{D.}~\bibnamefont{{Zhou}}},
  \bibinfo{author}{\bibfnamefont{J.}~\bibnamefont{{Sun}}},
  \bibinfo{author}{\bibfnamefont{D.}~\bibnamefont{{Akinwande}}},
  \bibnamefont{et~al.}, \bibinfo{journal}{ArXiv e-prints}
  (\bibinfo{year}{2015}), \eprint{1512.00604}.

\bibitem[{\citenamefont{{Zhou} et~al.}(2016)\citenamefont{{Zhou}, {Wu}, {Ning},
  {Li}, {Du}, {Chen}, {Zhang}, {Chi}, {Wang}, {Zhu} et~al.}}]{Zhou2016}
\bibinfo{author}{\bibfnamefont{Y.}~\bibnamefont{{Zhou}}},
  \bibinfo{author}{\bibfnamefont{J.}~\bibnamefont{{Wu}}},
  \bibinfo{author}{\bibfnamefont{W.}~\bibnamefont{{Ning}}},
  \bibinfo{author}{\bibfnamefont{N.}~\bibnamefont{{Li}}},
  \bibinfo{author}{\bibfnamefont{Y.}~\bibnamefont{{Du}}},
  \bibinfo{author}{\bibfnamefont{X.}~\bibnamefont{{Chen}}},
  \bibinfo{author}{\bibfnamefont{R.}~\bibnamefont{{Zhang}}},
  \bibinfo{author}{\bibfnamefont{Z.}~\bibnamefont{{Chi}}},
  \bibinfo{author}{\bibfnamefont{X.}~\bibnamefont{{Wang}}},
  \bibinfo{author}{\bibfnamefont{X.}~\bibnamefont{{Zhu}}},
  \bibnamefont{et~al.}, \bibinfo{journal}{Proceedings of the National Academy
  of Science} \textbf{\bibinfo{volume}{113}}, \bibinfo{pages}{2904}
  (\bibinfo{year}{2016}).

\bibitem[{\citenamefont{Roman-Perez and Soler}(2009)}]{Roman-Perez2009}
\bibinfo{author}{\bibfnamefont{G.}~\bibnamefont{Roman-Perez}} \bibnamefont{and}
  \bibinfo{author}{\bibfnamefont{J.~M.} \bibnamefont{Soler}},
  \bibinfo{journal}{Phys.~Rev.~Lett.} \textbf{\bibinfo{volume}{103}},
  \bibinfo{pages}{096102} (\bibinfo{year}{2009}).

\bibitem[{\citenamefont{Klimes et~al.}(2011)\citenamefont{Klimes, Bowler, and
  Michaelides}}]{Klimes2011}
\bibinfo{author}{\bibfnamefont{J.}~\bibnamefont{Klimes}},
  \bibinfo{author}{\bibfnamefont{D.~R.} \bibnamefont{Bowler}},
  \bibnamefont{and}
  \bibinfo{author}{\bibfnamefont{A.}~\bibnamefont{Michaelides}},
  \bibinfo{journal}{Phys.~Rev.~B} \textbf{\bibinfo{volume}{83}},
  \bibinfo{pages}{195131} (\bibinfo{year}{2011}).

\bibitem[{\citenamefont{Thonhauser et~al.}(2007)\citenamefont{Thonhauser,
  Cooper, Li, Puzder, Hyldgaard, and Langreth}}]{Thonhauser2007}
\bibinfo{author}{\bibfnamefont{T.}~\bibnamefont{Thonhauser}},
  \bibinfo{author}{\bibfnamefont{V.~R.} \bibnamefont{Cooper}},
  \bibinfo{author}{\bibfnamefont{S.}~\bibnamefont{Li}},
  \bibinfo{author}{\bibfnamefont{A.}~\bibnamefont{Puzder}},
  \bibinfo{author}{\bibfnamefont{P.}~\bibnamefont{Hyldgaard}},
  \bibnamefont{and} \bibinfo{author}{\bibfnamefont{D.~C.}
  \bibnamefont{Langreth}}, \bibinfo{journal}{Phys.~Rev.~B}
  \textbf{\bibinfo{volume}{76}}, \bibinfo{pages}{125112}
  (\bibinfo{year}{2007}).

\bibitem[{\citenamefont{Dion et~al.}(2004)\citenamefont{Dion, Rydberg,
  Schroder, Langreth, and Lundqvist}}]{Dion2004}
\bibinfo{author}{\bibfnamefont{M.}~\bibnamefont{Dion}},
  \bibinfo{author}{\bibfnamefont{H.}~\bibnamefont{Rydberg}},
  \bibinfo{author}{\bibfnamefont{E.}~\bibnamefont{Schroder}},
  \bibinfo{author}{\bibfnamefont{D.~C.} \bibnamefont{Langreth}},
  \bibnamefont{and} \bibinfo{author}{\bibfnamefont{B.~I.}
  \bibnamefont{Lundqvist}}, \bibinfo{journal}{Phys.~Rev.~Lett.}
  \textbf{\bibinfo{volume}{92}}, \bibinfo{pages}{246401}
  (\bibinfo{year}{2004}).

\bibitem[{\citenamefont{Vegard}(1930)}]{Vegard1930}
\bibinfo{author}{\bibfnamefont{L.}~\bibnamefont{Vegard}},
  \bibinfo{journal}{Zeitschrift Fur Physik} \textbf{\bibinfo{volume}{61}},
  \bibinfo{pages}{185} (\bibinfo{year}{1930}).

\bibitem[{\citenamefont{Simon and Peters}(1980)}]{SIMON1980}
\bibinfo{author}{\bibfnamefont{A.}~\bibnamefont{Simon}} \bibnamefont{and}
  \bibinfo{author}{\bibfnamefont{K.}~\bibnamefont{Peters}},
  \bibinfo{journal}{Acta Crystallographica Section B-structural Science}
  \textbf{\bibinfo{volume}{36}}, \bibinfo{pages}{2750} (\bibinfo{year}{1980}).

\bibitem[{\citenamefont{Nixon et~al.}(1966)\citenamefont{Nixon, Parry, and
  Ubbelohde}}]{NIXON1966}
\bibinfo{author}{\bibfnamefont{D.~E.} \bibnamefont{Nixon}},
  \bibinfo{author}{\bibfnamefont{G.~S.} \bibnamefont{Parry}}, \bibnamefont{and}
  \bibinfo{author}{\bibfnamefont{A.~R.} \bibnamefont{Ubbelohde}},
  \bibinfo{journal}{Proceedings of the Royal Society of London Series
  A-mathematical and Physical Sciences} \textbf{\bibinfo{volume}{291}},
  \bibinfo{pages}{324} (\bibinfo{year}{1966}).

\bibitem[{\citenamefont{Bronsema et~al.}(1986)\citenamefont{Bronsema, Deboer,
  and Jellinek}}]{BRONSEMA1986}
\bibinfo{author}{\bibfnamefont{K.~D.} \bibnamefont{Bronsema}},
  \bibinfo{author}{\bibfnamefont{J.~L.} \bibnamefont{Deboer}},
  \bibnamefont{and} \bibinfo{author}{\bibfnamefont{F.}~\bibnamefont{Jellinek}},
  \bibinfo{journal}{Zeitschrift Fur Anorganische Und Allgemeine Chemie}
  \textbf{\bibinfo{volume}{541}}, \bibinfo{pages}{15} (\bibinfo{year}{1986}).

\bibitem[{\citenamefont{Lee et~al.}(2010)\citenamefont{Lee, Murray, Kong,
  Lundqvist, and Langreth}}]{Lee2010}
\bibinfo{author}{\bibfnamefont{K.}~\bibnamefont{Lee}},
  \bibinfo{author}{\bibfnamefont{E.~D.} \bibnamefont{Murray}},
  \bibinfo{author}{\bibfnamefont{L.}~\bibnamefont{Kong}},
  \bibinfo{author}{\bibfnamefont{B.~I.} \bibnamefont{Lundqvist}},
  \bibnamefont{and} \bibinfo{author}{\bibfnamefont{D.~C.}
  \bibnamefont{Langreth}}, \bibinfo{journal}{Phys.~Rev.~B}
  \textbf{\bibinfo{volume}{82}}, \bibinfo{pages}{081101}
  (\bibinfo{year}{2010}).

\bibitem[{\citenamefont{Perdew et~al.}(1996)\citenamefont{Perdew, Burke, and
  Ernzerhof}}]{Perdew1996}
\bibinfo{author}{\bibfnamefont{J.~P.} \bibnamefont{Perdew}},
  \bibinfo{author}{\bibfnamefont{K.}~\bibnamefont{Burke}}, \bibnamefont{and}
  \bibinfo{author}{\bibfnamefont{M.}~\bibnamefont{Ernzerhof}},
  \bibinfo{journal}{Phys.~Rev.~Lett.} \textbf{\bibinfo{volume}{77}},
  \bibinfo{pages}{3865} (\bibinfo{year}{1996}).

\bibitem[{SM()}]{SM}
\bibinfo{note}{See Supplemental Material at
  http://link.aps.org/supplemental/xxx for other predicted
  structures, calculated lattice constants, phonon dispertion and molecular
  dynamic simulation of Pna21-CO}.

\bibitem[{\citenamefont{Parrinello and Rahman}(1980)}]{PARRINELLO1980}
\bibinfo{author}{\bibfnamefont{M.}~\bibnamefont{Parrinello}} \bibnamefont{and}
  \bibinfo{author}{\bibfnamefont{A.}~\bibnamefont{Rahman}},
  \bibinfo{journal}{Phys.~Rev.~Lett.} \textbf{\bibinfo{volume}{45}},
  \bibinfo{pages}{1196} (\bibinfo{year}{1980}).

\bibitem[{\citenamefont{Parrinello and Rahman}(1981)}]{PARRINELLO1981}
\bibinfo{author}{\bibfnamefont{M.}~\bibnamefont{Parrinello}} \bibnamefont{and}
  \bibinfo{author}{\bibfnamefont{A.}~\bibnamefont{Rahman}},
  \bibinfo{journal}{J. Appl. Phys.} \textbf{\bibinfo{volume}{52}},
  \bibinfo{pages}{7182} (\bibinfo{year}{1981}).

\end{thebibliography}

%\end{document}

\end{document}